\documentclass
[prd,twocolumn,showpacs,preprintnumbers]{revtex4}
\usepackage{amssymb}
\usepackage{graphicx}
\usepackage{dcolumn}
\usepackage{bm}
\usepackage{amsfonts}
\usepackage{latexsym}
\usepackage{amsmath}
\usepackage{psfrag}
\usepackage[cp1251]{inputenc}
\usepackage[dvips]{epsfig}

\newcommand{\lb}{\label}
\newcommand{\be}{\begin{equation}}
\newcommand{\ee}{\end{equation}}
\newcommand{\bea}{\begin{eqnarray}}
\newcommand{\eea}{\end{eqnarray}}
\newcommand{\bw}{\begin{widetext}}
\newcommand{\ew}{\end{widetext}}
\newcommand{\dv}{{\dot v}}
\newcommand{\da}{{\dot a}}
\newcommand{\dda}{{\ddot a}}
\newcommand{\ep}{{\varepsilon}}
\newcommand{\ffi}{{\varphi}}

\newcommand{\e}{{\rm e}}
\newcommand{\bn}{{\rm bound}}
\newcommand{\emi}{{\rm emit}}
\newcommand{\bi}{\bibitem}

\begin{document}
\preprint{DTP-MSU/04-06} \preprint{hep-th/0405121}

\title{Radiation reaction reexamined: bound momentum and Schott
term}

\author{ Dmitri V. Gal'tsov$^{*\dag}$\thanks{Email:  galtsov@grg.phys.msu.ru} and Pavel
Spirin$^*$} \affiliation{$^*$Department of Theoretical Physics,
Moscow State University, 119899, Moscow, Russia \\ $\dag$
Laboratoire de Math\'ematiques et Physique Th\'eorique, CNRS-UMR
6083, Universit\'e de Tours, Parc de Grandmont, 37200 Tours,
FRANCE}

\begin{abstract}
We review and compare two different approaches to radiation
reaction in classical electrodynamics of point charges: a local
calculation of the self-force using the charge equation of motion
and a global calculation consisting in  integration of the
electromagnetic energy-momentum flux through a hypersurface
encircling the world-line. Both approaches are complementary and,
being combined together, give rise to an identity relating the
locally and globally computed forces. From this identity it
follows that the Schott terms in the Abraham force should arise
from the bound field momentum and can not be introduced by hand as
an additional term in the mechanical momentum of an accelerated
charge. This is in perfect agreement with the results of Dirac and
Teitelboim, but disagrees with the recent calculation of the bound
momentum in the retarded coordinates. We perform an independent
calculation of the bound electromagnetic momentum  and verify
explicitly that the Schott term is the derivative of the finite
part of the bound momentum indeed. The failure to obtain the same
result using the method of retarded coordinates tentatively lies
in an inappropriate choice of the integration surface. We also
discuss the definition of the delta-function on the semi-axis
involved in the local calculation of the radiation reaction force
and demonstrate inconsistency of one recent proposal.

\end{abstract}
 \maketitle

\section{Introduction and overview}

Studies of the radiation reaction in classical electrodynamics
initiated by Lorentz and Abraham  as far as in the 19-th century,
remained an area of active research during the whole 20-th
century. Although with the development of quantum electrodynamics
this problem became somewhat academic, it still attracts attention
in connection with new applications and new ideas in fundamental
theory. The current understanding of the radiation reaction has
emerged in the classical works of Dirac~\cite{Di38}, Ivanenko and
Sokolov~\cite{IvSo48}, Rohrlich~\cite{Ro65},
Teitelboim~\cite{Te70} and some others. One of the crucial points
of this theory is the nature of the non-dissipative term in the
reaction force known as the Schott term~\cite{Sch15}. According to
Dirac~\cite{Di38} and Teitelboim~\cite{Te70}, this term is a
finite part of the derivative of the momentum of the
electromagnetic field which is bound to the charge. Meanwhile, as
far as we are aware, in the existing literature there are only two
explicit calculations of the bound momentum via integration of the
corresponding flux: that by Dirac and a more recent calculation by
Poisson~\cite{Po99}, who used the method of retarded coordinates
of Newman and Unti~\cite{NeUn63}. The results disagree, namely,
according to \cite{Po99}, the bound momentum produces only an
infinite term which has to be absorbed by the mass
renormalization.  The purpose of this paper is to clarify the
nature of the  above discrepancy and to provide an independent
calculation of the bound momentum.

In the famous paper published in 1938, Dirac \cite{Di38} gave a
detailed and consistent derivation of the equation of motion of a
point radiating charge in four-dimensional flat space-time. In his
original formulation, the decomposition of the retarded field into
the sum of the radiation field (the half-difference of the
retarded and advanced fields) and the self-action field (the
half-sum of the same quantities) was suggested. When such a
decomposition is used to describe the action of the proper field
on the charge itself, the first part provides the finite
radiation-reaction force (the Abraham vector~\cite{Ab05})
\be\lb{Ab} f^\mu_{\rm rad}=\frac23e^2(a^2v^\mu + \da^\mu),\ee
where $v^\mu$ is the four-velocity and $a^\mu =\dv^\mu$ is the
four-acceleration. Here the dot denotes the derivative with
respect to the proper time, $a^2=a_\mu a^\mu$, and the signature
$+,-,-,-$ is understood. The second part contributes only to an
infinite renormalization of mass and does not affect the motion of
the charge in the flat space-time.

The reaction force contains the radiation recoil term, the first
term in (\ref{Ab}), \be f^\mu_{\rm emit}=\frac23e^2a^2v^\mu, \ee
and the  Schott term \be f^\mu_{\rm Schott}=\frac23e^2 \da^\mu.
\ee  This second term is a total derivative, so it does not
correspond to an irreversible loss of momentum by the particle,
but  plays an important role in  the momentum balance between the
radiation and  particle momentum loss. If $f^\mu_{\rm rad}=0$,
this does not necessarily mean that there is no radiation (recall
the well-known case of radiation from a uniformly accelerated
charge \cite{IvSo48,FuRo60}), but if there is no radiation, the
Abraham force (\ref{Ab}) is zero. Indeed, if one has $a_\mu
a^\mu=0$ at any time, then it is easy to show that the
three-acceleration is zero, ${\bf a}=0$, and therefore $\dot a^\mu
=0$. Thus, no radiation reaction force is possible in the absence
of radiation.

The correct interpretation and a constructive derivation of the
Schott term was given by Dirac himself~\cite{Di38} via integration
of the Maxwell stress-tensor of the charge retarded field over the
space-like hypersurface orthogonal to the world-line. The
subsequent discussion was somehow obscured by the use in this
paper of the advanced potential. But Havas~\cite{Ha48} noticed
that actually only the physical retarded field was involved in the
calculation of the bound electromagnetic momentum, and the whole
expression for the Abraham vector can be obtained using only the
retarded field~\cite{Ro61}. This became especially transparent
after a later investigation  of the nature of the Schott term by
Teitelboim \cite{Te70} (see also the review \cite{TeViVa80}),
where it was emphasized that this term originates from the bound
electromagnetic momentum. However, having provided a very clear
and comprehensive discussion, Teitelboim did not present an
explicit calculation of the  quantities involved, addressing the
reader to the Dirac's paper for technical details. Meanwhile, the
details of the integration carried out in \cite{Di38} are fairly
non-trivial, and that is why more recently  an attempt was made in
Ref.~\cite{Po99} to simplify the  derivation using the retarded
coordinates of Newman and Unti \cite{NeUn63}. This simplified
method seems to be getting popular, and it has been generalized to
arbitrary dimensions~\cite{Ya03}. However, this modified
calculation fails to give the Schott term as a part of the bound
electromagnetic momentum. This has led the author of~\cite{Po99}
to revive the attempt to ascribe a {\em mechanical} origin to the
Schott term, which  interpretation can in fact be found in the
earlier literature.

The crucial point in  Dirac's calculation was the power series
expansion of the retarded field  in terms of the suitably defined
small parameter related to the proper time difference between the
moments of the emission and observation. This is necessary because
the integrand of the associated integral expression contains
retarded fields taken at different moments of the proper time.
Although one might think that such expansions can be avoided by
using the retarded coordinated, a more careful analysis shows that
this is not so. We present here a straightforward calculation
which is technically slightly different from that  used by Dirac,
but is similar conceptually. We have also generalized the
calculation to arbitrary space-time dimensions (to be presented
elsewhere), which sheds new light to the problem of the Schott
term. Note that multidimensional generalization of the Maxwell
theory was discussed by Ivanenko and Sokolov \cite{IvSo40}, soon
after the work of Dirac, in connection with the Huygens principle,
and recently this problem has attracted attention in view of
general interest to space-time models with large extra dimensions
\cite{Ko99,Ga01,Ka02,Ka03,Ya03}).

Some confusion about the Schott term is also related to the
well-known 'phenomenological' derivation of the Lorentz-Dirac
equation, as given in the book by Landau and Lifshitz \cite{LaLi}.
In this heuristic derivation, the first term of the Abraham force
is obtained by computing the rate of radiation, while the Schott
term is added by hand from the requirement of orthogonality of the
reaction force to the particle four-velocity. Formally, this
procedure leads to the correct equation (though does not answer
the question about the physical origin of the Schott term), so
{\em per se} it does not contradict to the correct interpretation
according to which the Schott term is viewed as a part of the
bound momentum. But as we show here, if one does not relate the
Schott term  to the electromagnetic momentum, the energy-momentum
balance equations become contradictory (Sect. II). The essential
difference between the 'phenomenological' derivation of the Schott
term via the orthogonalization procedure and its consistent
treatment as the derivative of the finite part of the bound
electromagnetic momentum becomes especially  clear in higher
dimensions. It turns out that generically the number of possible
momentum 'counterterms' in higher dimensions is larger than the
number of equations arising from the requirement of the
orthogonality. As a result, the Schott term(s) can not be obtained
within the orthogonalization procedure  in even dimensions higher
than six.

The redefinition of the mechanical momentum of a radiating charge
would also be wrong  conceptually, since it would imply that the
Maxwell-Lorentz electrodynamics {\em be not} muliplicatively
renormalizable:  a new (finite) mechanical term  not present in
the initial lagrangian would be required. In fact, failure of the
multiplicative renormalizability is what happens indeed in
higher-dimensional electrodynamics \cite{Ga01,Ka02}, but an
essential property of the {\em four-dimensional} theory is the
renormalizability in the sense that no  additional counterterms
(either  infinite or finite) are required to make the theory
consistent. This reflects the renormalizability of {\em quantum}
electrodynamics in four dimensions.

Our other remark concerns the regularization of  products of the
delta-function and its derivatives with the Heaviside function
which arise in the quasilocal treatment of the radiation reaction
problem. Recently this problem was reconsidered in \cite{Ka02} in
the context of the higher-dimensional generalization of the
Maxwell electrodynamics, where a new regularization scheme was
suggested and used to derive the Lorentz-Dirac equation in even
dimensions higher than four. We will show  below that the
regularization proposed in \cite{Ka02} fails to reproduce the
correct result already in the case of four dimensions, so the
validity of the equations derived in \cite{Ka02,Ka03} (and  of the
regularization itself) is questionable. Meanwhile the consistent
treatment of the delta-function on a half-line via the
point-splitting was suggested long ago (see e.g. the books by
Ivanenko and Sokolov~\cite{IvSo48},  Rohrlich~\cite{Ro65} and
Barut~\cite{Ba}), and we have checked that it works perfectly well
in any even space-time dimensions. The actual check of validity of
any regularization of this kind, as   was emphasized by Dirac,
consists in an alternative calculation of the reaction force by
integrating the rate of the variation of the bound electromagnetic
momentum.

\section{Field decomposition versus stress tensor decomposition}
Our definitions closely follow  those by Rohrlich~\cite{Ro65} and
Teitelboim~\cite{Te70}. The retarded potential generated by a
point charge moving along the world-line $x^\mu=z^\mu(s)$ depends
on the kinematic variables taken at the (position dependent)
retarded proper time $s_{\rm ret}(x)$ defined as the solution to
the equation \be R^{\mu} R_{\mu}=0,\quad
R^{\mu}=x^{\mu}-z^{\mu}(s_{\rm ret}),\ee satisfying $x^0>z^0$. The
second solution to the same equation with $z^0>x^0$ defines the
advanced proper time $s_{\rm adv}(x)$. Introducing the invariant
distance \be\lb{ro}\rho=v_{\mu}(s_{\rm ret})R^{\mu},\quad
v^\mu=\frac{dz^\mu}{ds}, \ee which is equal to the spatial
distance $|\mathbf{R}|=|\mathbf{x}-\mathbf{z}(s_{\rm ret})|$
between the points of emission and  observation  in the
momentarily co-moving Lorenz frame at the time moment
$x^0=z^0(s_{\rm ret})$, one can present the retarded potential as
\be\lb{wiechert4}
    A^{\mu}_{\rm ret}(x)=\frac{e v^{\mu}}{\rho}\Big|_{s_{\rm
    ret}(x)}.
\ee It is convenient to introduce the null vector
$c^{\mu}=R^{\mu}/\rho$, whose scalar product with $v^\mu$ is equal
to unity, and also the unit space-like vector
$u^\mu=c^{\mu}-v^{\mu}$. Thus we have
 \begin{align}\label{orts}
v^2=1, \quad c^2=0, \quad vc=1,\quad u^2 =-1.
\end{align}
(Here and below we omit, where unambiguous, brackets in the
four-dimensional scalar products, e.g. $vc=v^\mu c_\mu)$.
Differentiations with respect to $x^\mu$ are performed using the
relations
\begin{align}\label{difs}
&\partial_{\mu} s_{\rm
    ret}(x) =c_{\mu},\nonumber\\
&\partial_{\mu} \rho =v_{\mu}+ \lambda c_{\mu}, \\
&\partial_{\mu} c^{\nu}=\frac{1}{\rho}\left(\delta _{\mu}^{\nu}-
v_{\mu}c^{\nu}-c_{\mu}v^{\nu}-\lambda c_{\mu}c^{\nu}, \nonumber
 \right)
\end{align}
where \be \lambda=\dot{\rho}=\rho (ac)-1.\ee  Using these formulas
we obtain the field tensor
\begin{equation}\label{fieldtensor4}
 F_{\mu
 \nu}=\frac{e \left(\rho(ac)-1\right)}{\rho^2}v_{[\mu}c_{\nu]}-\frac{e }{\rho}a_{[\mu}c_{\nu]}
\end{equation}
where square brackets denote antisymmetrization without factor
$1/2$, {\em e.g.}
$a_{[\mu}c_{\nu]}=a_{\mu}c_{\nu}-a_{\nu}c_{\mu}$, and all
quantities have to be taken at the moment $s_{\rm ret}(x)$ of the
proper time . Similarly, the advanced potential $A^\mu_{{\rm
adv}}$ and the corresponding field can be written in terms of
quantities depending on $s_{\rm adv}(x)$.

Separation of the radiation from the total field can be performed
in two different ways. The first procedure is linear in  the field
and consists in the splitting the retarded potential $A^\mu_{{\rm
ret}}$ into the radiative part \be\lb{Arad} A^\mu_{{\rm
rad}}=\frac12\left(A^\mu_{{\rm ret}}-A^\mu_{{\rm adv}}\right) \ee
and the 'self' part \be\lb{Aself} A^\mu_{{\rm
self}}=\frac12\left(A^\mu_{{\rm ret}}+A^\mu_{{\rm adv}}\right).
\ee The radiative potential satisfies the homogeneous D'Alembert
equation and changes the sign under reflection of time, as
expected for the radiation irreversibly lost by an accelerated
charge. This part of the retarded field tends to zero in the
static limit. The self part is time-symmetric and remains finite
in the static limit, where it coincides with the Coulomb
potential.

The split of the second kind  is quadratic in  the field and uses
the energy-momentum tensor \be\lb{maxstress}
T^{\mu\nu}=\frac1{4\pi}\left(F^{\mu\lambda}F_{\lambda}^{\;\nu}
+\frac14\eta^{\mu\nu}F^{\alpha\beta}F_{\alpha\beta}\right).\ee
Considering the energy-momentum tensor we will always deal with
the retarded solution of the D'Alembert equation. Substituting
here the field tensor (\ref{fieldtensor4}) we obtain the sum of
two terms \be \lb{stressplit}T^{\mu\nu}=T^{\mu\nu}_{\rm
emit}+T^{\mu\nu}_{\rm bound},\ee where the first term is
proportional to $\rho^{-2}$:
\begin{equation}\label{SET4Dradiat}
\frac{4\pi}{e^2} T^{\mu\nu}_{\rm emit}=-
\frac{(ac)^2+a^2}{\rho^2}c^{\mu}c^{\nu},
\end{equation}
while the second one contains higher powers of
$\rho^{-1}$:\bea\label{Tbound4c} \frac{4\pi}{e^2} T^{\mu \nu}_{\rm
 bound}&=&\frac{a^{(\mu}c^{\nu)}+2(ac)c^{\mu}c^{\nu}-
 (ac)v^{(\mu}c^{\nu)}}{\rho^3} \nonumber \\&+&
 \frac{v^{(\mu}c^{\nu)}-c^{\mu}c^{\nu}- \eta^{\mu
 \nu}/2}{\rho^4}.
\eea  Here the symbol $(\mu\nu)$ denotes a symmetrization without
the factor $1/2$.

The first expression (\ref{SET4Dradiat}) is distinguished by the
following properties:
\begin{itemize}\item its geometric structure is the tensor product
of two null vectors $c^\mu$, \item it is traceless, \item it falls
down as $|{\mathbf x}|^{-2}$ when $|{\mathbf x}|\to \infty$, \item
it is independently conserved
\begin{equation}\label{radconserv}
\partial_{\nu}T^{\mu \nu}_{\rm emit}=0.
\end{equation}
\end{itemize}
This latter property follows from the differentiation rules
(\ref{difs}). All these features indicate that $T^{\mu\nu}_{\rm
emit}$ describes an outgoing radiation. An independent
conservation of this quantity means that the bound part is also
independently conserved
\begin{equation}\label{boundconserv}
\partial_{\nu}T^{\mu \nu}_{\rm bound}=0.
\end{equation}

Conservation of the total four-momentum implies that the sum of
the mechanical momentum and the momentum of the electromagnetic
field is constant (for simplicity we do not include an external
field ): \be \frac{d p^\mu_{\rm mech}}{ds}+\frac{d p^\mu_{\rm
em}}{ds}=0. \ee Here the mechanical part is proportional to the
bare mass of the charge \be p^\mu_{\rm mech}=m_0 v^\mu,\ee while
the field part is given by \be p^\mu_{\rm em} =\int
T^{\mu\nu}d\Sigma_\nu,\ee where integration of the electromagnetic
stress tensor is performed over a space-like hypersurface whose
choice will be discussed in detail later on. It has to be
emphasized again that in the expression (\ref{maxstress}) for the
stress tensor of the electromagnetic field one has to use the
physical retarded field. According to the second splitting, one
can write \bea \frac{d p^\mu_{\rm mech}}{ds}&=&-\frac{d
p^\mu_{\rm em}}{ds}=f^\mu_{\rm emit}+f^\mu_{\rm bound},\lb{one}\\
f^\mu_{\rm emit}&=&-\frac{d}{ds}\int T^{\mu\nu}_{\rm emit}d\Sigma_\nu,  \\
f^\mu_{\rm bound}&=&-\frac{d}{ds}\int T^{\mu\nu}_{\rm
bound}d\Sigma_\nu.\eea

On the other hand, the derivative of the bare mechanical momentum
can be expressed using the  equation of motion of the charge in
which the electromagnetic field is decomposed into the self
part~(\ref{Aself}) and the radiation part~(\ref{Arad}) \bea
\frac{d p^\mu_{\rm mech}}{ds}&=&e F^{\mu\nu}_{\rm ret}v_\nu
\nonumber\\&=&e\left( F^{\mu\nu}_{\rm self}+F^{\mu\nu}_{\rm
rad}\right)v_\nu \nonumber
\\ &=& f^\mu_{\rm self}+f^\mu_{\rm rad}.\lb{two}\eea Clearly, the
following energy-momentum conservation identity should hold in
view of (\ref{one}) and (\ref{two}):\be \lb{consistency}f^\mu_{\rm
self}+f^\mu_{\rm rad}=f^\mu_{\rm bound}+f^\mu_{\rm emit}.\ee

Now, somewhat unexpectedly, $ f^\mu_{\rm rad}\neq f^\mu_{\rm
emit}$ and $ f^\mu_{\rm self}\neq f^\mu_{\rm bound}$, differing by
the Schott term:\bea f^\mu_{\rm rad}&=& f^\mu_{\rm emit}+
f^\mu_{\rm Schott},\lb{rad-emit}\\ f^\mu_{\rm self}&=& f^\mu_{\rm
bound}- f^\mu_{\rm Schott}.\lb{Schott}\eea The identity
(\ref{consistency}) is satisfied as expected. Explicit
calculations verifying these results will be presented in what
follows. They are fully consistent with the results of Dirac
\cite{Di38}, Rohrlich \cite{Ro65} and Teitelboim \cite{Te70}. On
the contrary, in \cite{Po99} it was found that $f^\mu_{\rm bound}
=f^\mu_{\rm self}$, while the Eq. (\ref{rad-emit}) still holds.
This is obviously inconsistent with the energy-momentum
conservation identity  (\ref{consistency}).

To avoid any confusion, we note that though  both sides of the
energy-momentum conservation identity  (\ref{consistency}) contain
divergent terms, the extraction of  finite terms  is fully
unambiguous because of their different dependence on the
kinematical variables. Moreover, the parameterization of the
divergent terms can be made similar in both calculations, so these
terms mutually cancel in (\ref{consistency}) before the
regularization is removed.
\section{World-line calculation: point splitting}
The retarded and advanced potential taken on the world-line
$x^\mu=z^\mu(s)$ of a charge can conveniently be written in terms
of  Green's functions \cite{IvSo48} \bea
\lb{Gsr} G_{\rm self}(Z)=\delta{( Z^2)}, \\
G_{\rm rad}(Z)=\frac{Z^0}{|Z^0|}\delta{( Z^2)}, \eea where
$Z^\mu=Z^\mu(s,s')=z^\mu(s)-z^\mu(s')$. Substitution  of the
electromagnetic field of the  charge on its world-line leads to
the following integrals   \be f^\mu(s)=2
e^2\int\;Z^{[\mu}(s,s')v^{\nu]}(s')v_\nu(s)\frac{d}{d Z^2} G(Z)
ds', \ee  for both $f^\mu_{\rm self}$ and $f^\mu_{\rm rad}$. Due
to  the presence of delta-functions in  $G_{\rm self}$ and $G_{\rm
rad}$, one is tempted to expand the integrands in $\sigma=s-s'$.
Taking into account that $Z^2=\sigma^2+O(\sigma^4)$, one can write
\bea\lb{sings} G_{\rm self}(Z)&=&
\delta(\sigma^2)+O(\sigma^4),\\\lb{sings1} G_{\rm rad}(Z)&=&
\frac{\sigma}{|\sigma|} \left(\delta(\sigma^2)+O(\sigma^4)\right).
\eea Expanding the rest of the integrands in $\sigma$, one
encounters the following integrals:  \be\lb{A}
A_l=\int_{-\infty}^\infty \sigma^l\,\frac{d}{d
\sigma^2}\delta(\sigma^2)\;d\sigma, \ee for the self-force, and
\be \lb{B}B_l=\int_{-\infty}^\infty \sigma^l\,\frac{d}{d
\sigma^2}\left(
\frac{\sigma}{|\sigma|}\delta(\sigma^2)\right)\,d\sigma \ee for
the radiation reaction force, with $l\geq 2$.

Both these integrals are ill-defined.  Passing to the variable
$x=\sigma^2$ they can be transformed to integrals of the type \be
\int_0^\infty \phi(x) \delta^{(k)}(x)dx,\ee where
$\delta^{(k)}(x)$ denotes  the $k$-s derivative of the
delta-function, which function has the support at the boundary
point of the integration domain. Computing such an integral is
equivalent to taking the product of the delta-function and the
Heaviside function, or, equivalently, defining $\delta(x)$ on the
semi-axis. For this a suitable regularization is needed. Before
discussing this point, we note that,  with any regularization, the
integrals (\ref{A}), (\ref{B}) should vanish for $l>3$ by power
counting. Moreover, all $A_l$ vanish  for odd $l$, and all $B_l$
vanish  for even $l$ from parity considerations. By power counting
one can also show that terms in (\ref{sings}),~(\ref{sings1})
proportional to $\sigma^4$ actually give no contribution, while
the relevant terms that  must be retained in the expansion of the
integrand are given by \be\lb{exp}
2X^{[\mu}(s,s')v^{\nu]}(s')v_\nu(s)=a^\mu\sigma^2-
\frac23(\da^\mu+v^\mu a^2 )\sigma^3. \ee

Now we  discuss the meaning of the delta-function on the
semi-axis. Recently an attempt was made \cite{Ka02} to develop a
general theory for such objects using primary definitions of the
theory of distributions. The proposed regularization for the first
derivative of the delta-function defined on the semi-axis (see Eq.
(23) in \cite{Ka02}) reads: \be \delta'(x)=\lim_{\alpha\to +0}
\frac{\partial}{\partial
\alpha}\frac{\e^{-x/\alpha}}{\alpha},\quad x\geq 0.\ee Using this
regularization one finds: \bea A_2&=&\lim_{\alpha\to 0}
\frac{\sqrt{\pi}}{4\sqrt{\alpha}},\\ B_3&=& 1.\eea  Substituting
this into the above formulas we obtain \bea f^\mu_{\rm
self}&=&\lim_{\alpha\to 0} \frac{e^2\sqrt{\pi}}{4\sqrt{\alpha}} a^\mu,\\
f^\mu_{\rm rad}&=&-\frac{2e^2}3(v^\mu a^2+\da^\mu).\eea The first
quantity can be  absorbed by the renormalization of mass, while
the second gives the correct structure for the Abraham force but
with the {\em wrong sign}. Thus the proposal \cite{Ka02} for the
delta-function and its derivatives  on the semi-axis  fails to
give the correct result.

Meanwhile, a satisfactory method to deal with such integrals was
suggested long ago (see~\cite{IvSo48,Ro65,Ba}), it consists in the
'point-splitting'   \be \delta(\sigma^2)=\lim_{\ep\to +0}
\delta(\sigma^2-\ep^2)=\lim_{\ep\to
+0}\frac{\delta(\sigma-\ep)+\delta(\sigma+\ep)}{2\ep}\ee with a
prescription that the limit should be taken only after evaluating
all the integrals. With this, omitting the symbol of the limit, we
obtain \bea A_2&=-&  \frac1{2 \ep },\\
B_3&=& -1,\eea so one finds \bea \lb{fself} f^\mu_{\rm
self}&=&-  \frac{e^2 }{2 \ep} a^\mu,\\
f^\mu_{\rm rad}&=&\frac{2e^2}3(v^\mu a^2+\da^\mu).\eea After the
mass renormalization, \be m_0-A_2=m,\ee  we get the Lorentz-Dirac
equation \be m a^\mu= \frac{2e^2}3(v^\mu a^2+\da^\mu), \ee

One has to realize, however,  that there is no purely mathematical
proof of the correctness of the point splitting precedure. So
actually to check the validity of this or any other regularization
involved in the local calculation of the radiation reaction force
one has to perform an alternative calculation  via the integration
of the electromagnetic stress-tensor.

\section{Integration of  electromagnetic momentum}
\subsection{General setting}

An alternative derivation of the radiation reaction force consists
in the integration of the energy-momentum tensor of the
electromagnetic field of a charge over an appropriate hypersurface
in space-time. The correct choice of this hypersurface is
essential for the calculation, so we will discuss it in detail.
Assuming the split (\ref{stressplit}) of the stress-tensor into
the emitted and bound parts, one can consider the emitted and
bound momenta separately. Teitelboim \cite{Te70} defined the
corresponding integration surfaces differently in both cases
taking into account the specific properties of these quantities.
Here we will show that both integrals can be transformed to those
over an infinitely thin world tube around the particle world-line.

We would like to calculate the four-momentum carried by the
electromagnetic field of the charge  for a given moment of the
proper time $s$ on the particle world-line $z^\mu(s)$. To do this
one has to choose a space-like hypersurface $\Sigma (s)$
intersecting the world-line at the point $z^\mu(s)$ and to
integrate the electromagnetic energy-momentum flux  as follows
\begin{align}\label{momentdefinition}
p_{\rm em}^{\mu}(s)=\int \limits_{\Sigma(s)}T^{\mu
\nu}d\Sigma_{\nu}.
\end{align}
The simplest and the most practical choice for $\Sigma (s)$ is
that of the hypersurface orthogonal to the world-line
\be\label{hypersigma} v_\mu(s)\left(x^\mu-z^\mu(s)\right)=0. \ee
\begin{figure}
\centerline{\psfrag{s1}{\LARGE $ s_{1}$}    \psfrag{s2}{\LARGE $
s_{2}$}   \psfrag{Sie}{\LARGE $S_{\varepsilon}$}
\psfrag{Sis1}{\LARGE $ \Sigma(s_1)$}    \psfrag{dYRs1}{\LARGE $
\partial Y_R(s_1)$}
\psfrag{Yis2}{\LARGE $ Y(s_2)$}\psfrag{SiR}{\LARGE $ S_R$}
\psfrag{zmis}{\LARGE $ z^{\mu}(s)$}  \psfrag{dYes1}{\LARGE
$\partial Y_{\varepsilon}(s_1)$}
\includegraphics
[angle=270,width=15cm]{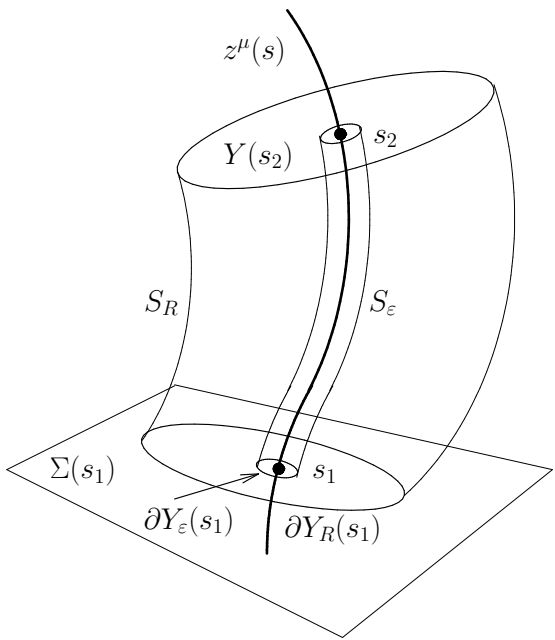}}\caption{Integration  of the
bound electromagnetic  momentum. Here  $\Sigma(s_1)$ is the
space-like  hyperplane transverse to the world-line  $z^\mu(s)$
intersecting it at the proper time $s_1$ (similarly
$\Sigma(s_2)$). The hypersurfaces $S_{\varepsilon}$ and $S_R$ are
small and large tubes around the world-line formed by sequences of
the 2-spheres $\partial Y_{\varepsilon}(s)$ and $\partial Y_R(s)$
for $s\in [s_1, \,s_2]$. The domain $Y(s_2)\subset \Sigma(s_2)$
(similarly $Y(s_1)$) is the 3-annulus between $\partial Y_R(s_2)$
and $\partial Y_{\varepsilon}(s_2)$.} \label{figur2}
\end{figure}
The integral (\ref{momentdefinition}) is divergent near the world
line. The rate of the divergence can be controlled introducing the
small length parameter $\ep$, the radius of the 2-sphere $\partial
Y_\ep(s)$ (Fig. 1), defined by the intersection of the hyperplane
(\ref{hypersigma}) with the hyperboloid \be\label{sphere}
(x-z(s))^2=-\varepsilon^2.\ee We also  introduce the sphere
$\partial Y_R(s)$ of a large radius $R$ defined by the
intersection of  $\Sigma(s)$ with the hyperboloid
\be\label{sphereR} (x-z(s))^2=- R^2.\ee The electromagnetic
momentum can then be obtained by taking the limit $\ep\to 0,\;
R\to \infty$ of the integral over the domain $Y(s)\subset
\Sigma(s)$ between the boundaries $\partial Y_\ep(s)$ and
$\partial Y_R(s)$.

Let us evaluate the variation of this quantity between the moments
$s_1$ and $s_2$ of the proper time on  the world-line of the
charge \be \Delta p^\mu_{\rm em}=\int \limits_{Y(s_2)}T^{\mu
\nu}d\Sigma_{\nu}-\int \limits_{Y(s_1)}T^{\mu
\nu}d\Sigma_{\nu}.\ee For the bound momentum it is convenient to
consider the tubes $S_\ep$ and $S_R$ formed as sequences of the
spheres $\partial Y_\ep(s)$ and $\partial Y_R(s)$ on the interval
$s\in [s_1,\,s_2]$ and to transform this quantity to \be \Delta
p^\mu_\bn=\int \limits_{S_R }T^{\mu \nu}_\bn dS_{\nu}-\int
\limits_{S_\ep}T^{\mu \nu}_\bn dS_{\nu}\ee in view of the
conservation equation for $T^{\mu\nu}_\bn $ (\ref{boundconserv}).
Here  normal vectors in $dS_\nu$ are directed outwards with
respect to the world-line. The contribution from the infinitely
distant surface $S_R$ vanishes if one assumes that the charge
acceleration is zero  in the limit $s\to-\infty$~\cite{Te70}. This
assertion is somewhat non-trivial, since, in spite of the fact
that the stress tensor (\ref{Tbound4c}) decays as $R^{-3}$, the
corresponding flux does not vanish {\em a priori}, because the
surface element contains a term (proportional to the acceleration)
which asymptotically grows as $R^3$ (see the Eq. \ref{arear}
below). As  a consequence, the surviving term will be proportional
to the acceleration taken at the moment $s_{\rm ret}$ of the
proper time, where $s_{\rm ret}\to -\infty$  in the limit
$R\to\infty$. Finally we are left with the integral over the inner
boundary only \be \lb{momentvardef1}\Delta
p^\mu_\bn=-\int\limits_{S_\ep} T^{\mu\nu}_\bn dS_\nu.\ee

For integration of the emitted momentum it is convenient to take
the light cone boundary $C(s')$ instead of $S_R$ as shown in
Fig.~2.
\begin{figure}
\centerline{\psfrag{s1}{\huge $s_{1}$}    \psfrag{s2}{\huge
$s_{2}$}   \psfrag{ss}{\huge $s'$} \psfrag{Css}{\huge $C(s')$}
\psfrag{Siis1}{\huge $\Sigma(s_1)$} \psfrag{Sisis1s2}{\huge
$C(s',s_1,s_2)$}\psfrag{Sie}{\huge $S_{\varepsilon}$}
\psfrag{zmis}{\huge $z^{\mu}(s)$}  \psfrag{Zis2}{\huge
$Z_{\varepsilon}(s',s_2)$} \psfrag{dZis2}{\huge $\partial
Z(s',s_2)$} \psfrag{dYeis2}{\huge $\partial Y_{\varepsilon}(s_2)$}
\includegraphics
[angle=270,width=11cm]{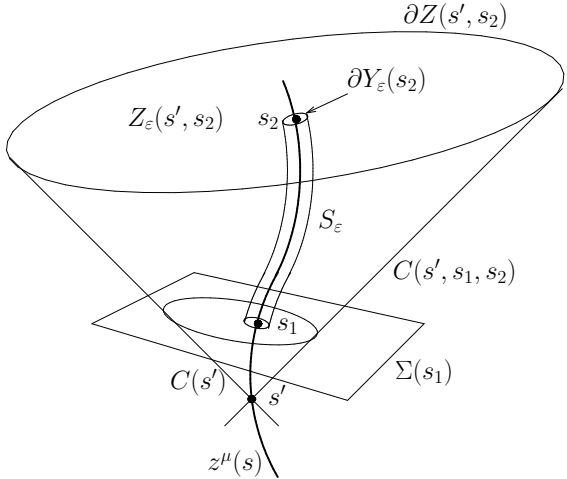}}\caption{Integration   of the
emitted  momentum. Here $C(s')$ is the future light cone of some
point $s'$ on the world-line, and $C(s',s_1,s_2)$ is its part
between the hypersurfaces $\Sigma(s_1)$ and $\Sigma(s_2)$. The
domain $Z(s',s_2)$ (similarly $Z(s',s_1))$ is the annulus between
the intersection of the light cone with $\Sigma(s_2)$ (the outer
boundary) and the small sphere $\partial Y_{\varepsilon}(s_2)$
(the inner boundary).} \label{figur4}
\end{figure}
The actual change of the emitted momentum in the whole three-space
corresponds to the limits $s'\to -\infty, \ep\to 0$. Since the
normal to the light cone lies on it, the flux of the
energy-momentum tensor through the null boundary $C(s',s_1,s_2)$
is zero for any $s'$, therefore
\begin{align}\label{emitmomentvardef1}
& \Delta p^{\mu}_{\rm emit} = \nonumber \\
&= \int\limits_{Z(s',s_2)} T^{\mu \nu}_{\rm emit} v_{\nu}(s_2)dS-
\int\limits_{Z(s',s_1)}
T^{\mu \nu}_{\rm emit} v_{\nu}(s_1)dS =  \nonumber  \\
& =-\int\limits_{S_\ep} T^{\mu \nu}_{\rm emit} dS_{\nu}+
\int\limits_{C(s',s_2,s_1)} T^{\mu
\nu}_{\rm emit} c_{\nu}dS=  \nonumber   \\
&= -\int\limits_{S_\ep} T^{\mu \nu}_{\rm emit} dS_{\nu}
\end{align}
where the limits $\ep \to 0$, $s'\to -\infty$ have to be taken. So
both the emitted and bound momenta can be reduced to  integrals
over the small tube around the world-line.

Now we have to find an integration measure on the small tube.
\subsection{Induced metric on $S_\ep $}  Consider the foliation of the space-time
region shown in Fig. 1 by the hypersurfaces $\Sigma(s)$
parameterized by the spherical coordinates $r, \theta_1=\theta,
\theta_2=\ffi$. Introducing the unit space-like vector $n^\mu
(s,\theta_i),\,n_\mu n^\mu=-1$, transverse to $v^\mu$, we can
write the following coordinate transformation from $x^\mu$ to the
set $s,r,\theta_i$:
\begin{align}\label{transferformulae}
& x^{\mu}=z^{\mu}(s)+r n^{\mu}(s,\theta_{i} ),\\&
v_{\mu}(s)n^{\mu}(s,\theta_{i})=0,\\ &
n_{\mu}(s,\theta_{i})n^{\mu}(s,\theta_{i})=-1.
\end{align}
 Since the 4-acceleration
vector $a^\mu$ is orthogonal to the 4-velocity, it lies in the
hyperplane $\Sigma(s)$ and we can further specify the angular
coordinates choosing the polar axis along the three-acceleration
${\bf a}$. Our convention about the angle variables is such that
the four-vector $n^\mu$ in the Lorentz frame momentarily comoving
with the charge at the proper time moment $s$,  CF(s), is given by
\begin{align}\label{n in MCLF}
n_c^\mu= (0,\rm \sin\theta\cos\ffi,sin\theta\sin\ffi, \cos\theta).
\end{align}
Then $a_\mu n^\mu=-a \cos \theta$, where $a=|{\bf a}|$.
Differentiating  (\ref{transferformulae}) with respect to the new
coordinates we obtain:
\begin{align}\label{transferderive}
&\frac{\partial x^{\mu}}{\partial s}=v^{\mu}(s)+r \frac{\partial
n^{\mu}(s,\theta_{i})}{\partial s},\\ &\frac{\partial
x^{\mu}}{\partial r}=n^{\mu}(s,\theta_{i}),\\& \frac{\partial
x^{\mu}}{\partial \theta_i}=r\frac{\partial
n^{\mu}(s,\theta_{i})}{\partial \theta_i}.
\end{align}

Let us calculate the derivative $\partial
n^{\mu}(s,\theta_{i})/\partial s$ in  CF($s$). To find the
variation $n^{\mu}(s+ds)- {n}_c^{\mu}$, we consider another
Lorentz frame, CF($s+ds$), which is comoving with the charge at
$s+ds$. These frames are   related by the Lorentz boost with the
velocity $|\mathbf{v}(s+ds)-\mathbf{v}(s)|=|\mathbf{a}(s)ds|=ads$.
Performing the Lorentz transformations and taking into account
that $\cos\theta$ is the same in both frames one finds
$n^\mu(s+ds)$ in CF($s$):
$$ n^\mu(s+ds)=\left(a{\rm cos }\theta
ds,\sin\theta\cos\ffi,\sin\theta\sin\ffi,{\rm cos} \theta\right).
$$ Hence in this frame
$$\frac{n^\mu(s+ds)-n^\mu(s)}{ds}=(a\cos \theta,0,0,0),$$
which is proportional to the four-velocity in the same frame
$v^\mu$=(1,0,0,0). Thus in an arbitrary frame one has
\begin{align}\label{n derive s}
\frac{\partial n^{\mu}}{\partial s}=a{\rm cos} \theta
v^{\mu}=-(an)v^{\mu}.
\end{align}

Denoting the angle derivatives as
$$e_i^{\mu}= r\partial n^{\mu}(s,\theta_{i})/\partial \theta_i,$$
we find the relations
\begin{align}\label{vierbein}
& n_{\mu}e_j^{\mu}=r\frac{\partial n^2}{2 \partial \theta_j}=0, \nonumber\\
& v_{\mu}e_j^{\mu}=r\frac{\partial (vn)}{\partial \theta_j}-r
n_{\mu}\frac{\partial v^{\mu}}{\partial \theta_j}=0, \\
& n_{\mu}v^{\mu}=0,\nonumber
\end{align}
showing that the four vectors $\{ n^{\mu},v^{\mu},e_j^{\mu}\}$
represent the space-time vierbein, with $e_j^{\mu}$ being the
tangent vectors to the 2-sphere. The induced metric on the sphere
is  $g_{ij}=\eta_{\mu\nu}e_i^\mu e_j^\nu$, and for the
four-dimensional induced metric  we obtain
\begin{align}\label{fullmetricsmatr}
g_{\mu\nu}=\left(
\begin{array}{ccc}
  [1-r(an)]^2 & 0 & 0 \\
  0 & -1 & 0 \\
  0 & 0 & -g_{ij} \\
\end{array}
\right),\end{align}\\
and therefore
\begin{align}\label{determetrics}
\det g=[1-r(an)]^2 \det g_{ij}.
\end{align}
Hence the  area element on the hypersurface $r=$const $S_\ep$ will
be given by
\begin{align}\label{arear}
d^{3}\sigma=r^2[1-r(an)]ds d\Omega,
\end{align}
where $d\Omega=\sin\theta d\theta d\ffi$.  Setting $r=\ep$ one
finds for the induced metric on $S_\ep$
\begin{align}\label{areaelement}
d^{3}\sigma=\varepsilon^{2}[1-\varepsilon(an)]ds d\Omega,
\end{align}
From~(\ref{vierbein}) one can see that $n^{\mu}$ is a unit vector
normal to the tube, so finally
\begin{align}\label{areaelement}
dS_{\mu}=\varepsilon^{2}[1-\varepsilon(an)]n_{\mu} ds d\Omega.
\end{align}
Substituting this  into
(\ref{emitmomentvardef1},\ref{momentvardef1}) we find for the
derivative of the electromagnetic momentum the following
expression \be\lb{dps}\frac{dp^\mu_{\rm em}}{ds}=-\int \ep^2
[1-\ep(an)]T_{\rm em}^{\mu\nu}n_\nu d\Omega,\ee valid for both the
emitted and bound parts, where the limit $\ep\to 0$ is understood.
It has to be realized that the dependence of $T_{\rm em}^{\mu\nu}$
on $\ep$ is somewhat non-trivial. In fact, the energy-momentum
tensor depends on the space-time point $x^\mu$ through the
quantity $\rho$, depending directly on $x^\mu$, and also through
the retarded proper time $s_{\rm ret}$. At the same time, we need
to express the resulting quantity as a function of the proper time
$s$ corresponding to the intersection of the world-line with the
space-like hypersurface. Carefully keeping track of all this we
expand the stress-tensor in terms of $\ep$ as follows
$$T_{\mu\nu}(\rho,s_{\rm{ret}})\big|_{S_\ep}=\sum_{k=-4}^\infty
{\varepsilon^{k}}
\Theta^k_{\mu\nu}(s,n^{\mu},v^{\mu},a^{\mu},\da^\mu).$$ The
leading divergent terms here are proportional to $\ep^{-2}$ for
the emitted part (\ref{SET4Dradiat},) and to $\ep^{-4}$ for the
bound part (\ref{Tbound4c}). Substituting this expansion into
(\ref{dps}) we have to perform integration over the angles and
then to take the limit $\ep\to 0$. The integration  is easily done
using the formula
\begin{align}  \label{sphereints}  & \int n_{\mu}n_{\nu}d\Omega
=\frac{4\pi}{3}\Delta_{\mu\nu},
 \quad \Delta_{\mu\nu}=v_{\mu} v_{\nu}-g_{\mu \nu},
\end{align} while the integration of products of an odd number of  $n_\mu$
gives zero.

\subsection{Emitted momentum} Considering first the emitted momentum case,
we see from  Eq. (\ref{SET4Dradiat}) that  the area factor
$\varepsilon^2$ compensates the denominator $\rho^2(\varepsilon)
\sim \varepsilon^2$ in the expression for $T^{\mu \nu}_{\rm
emit}$, so in the limit $\varepsilon \to 0$ it is sufficient to
take only the leading terms in the numerator
\begin{align}\label{emitexpans}
 a^2|_{s_{\rm ret}}& =a^2,\nonumber
\\c^{\mu}|_{s_{\rm ret}}& =n^{\mu}+v^{\mu},\\
(ac)^2|_{s_{\rm ret}}& =(an)^2,
\end{align}
where all quantities in the right hand sides are taken at the
proper time moment $s$. Omitting also the $\ep$-term in the
integration measure in ~(\ref{emitmomentvardef1}) and assuming
$s_1=-\infty,s_2=s$ we obtain
\begin{align}\label{emitflux}
p^{\mu}_{\rm emit}(s)=-\frac{e^2}{4\pi}\int
\limits_{-\infty}^{s}ds' \left(a^2+(an)^2\right)(v^{\mu}+n^{\mu})
d\Omega.
\end{align}
 After integration
over the angles  we arrive at
\begin{align}\label{emitflux1}
p^{\mu}_{\rm emit}(s)=-\frac{2e^2}{3}\int \limits_{-\infty}^{s}
ds' a^2 v^{\mu}(s'),
\end{align}
so the 'emitted' contribution to the radiative force is
\begin{align}\label{emitforce}
f^{\mu}_{\rm emit}=-\frac{dp^\mu_{\rm emit}}{ds}=\frac{2e^2}{3}
a^2 v^{\mu}.
\end{align}

\subsection{Bound momentum} In this case calculations are
substantially more involved. All quantities in~(\ref{Tbound4c})
depend on the retarded time, and to facilitate their expansion in
$\ep$-series it is useful to express $T_{\mu \nu}^{\rm bound}$
through the null vector $R^\mu=c^\mu\rho$:
\begin{align}\label{Tbound4R}
\frac{4\pi}{e^{2}} T^{\mu \nu}_{\rm bound}&= \frac
 {a^{(\mu}R^{\nu)}}{\rho^4}+\frac{\left(2(aR)-1\right)R^{\mu}R^{\nu}}{\rho^6}+
 \nonumber \\
 &+\frac{\left(1-(aR)\right)v^{(\mu}R^{\nu)}}{\rho^5}
  -\frac{\eta^{\mu \nu}}{2\rho^4}.
\end{align}
The expansion of $R^\mu$  reads
\begin{align}\label{R(sig)}
&
R^\mu=x^\mu-z^\mu(s_{\rm{ret}})=x^\mu-z^\mu(s)+z^\mu(s)-z^\mu(s_{\rm{ret}})=
\nonumber
\\ & =\varepsilon
n^\mu+v^\mu
 \sigma-\frac{1}{2}a^\mu \sigma^2+\frac{1}{6}\dot{a}^\mu \sigma^3
 +\mathcal{O}(\sigma^4),
\end{align}
where  $\sigma=s-s_{\rm{ret}}>0$ and all the vectors in the last
line are taken at $s$. This is an expansion in powers of $\sigma$,
but we need an expansion in powers of $\varepsilon$. The relation
between the two can be found  from the condition $R^2=0$. Assuming
an expansion of $\sigma$ in terms of $\ep$
\begin{align}\label{sigmaindcoeffs}
\sigma=\sum_{k}b_k\varepsilon^k,
\end{align}
and substituting it into the equation $R^2=0$, one finds order by
order the  coefficients:  $b_0=0, b_1=1, b_2=an/2 $ and so on.
Thus,  up to the third order terms, which is sufficient for our
purposes, we obtain:
\begin{align}\label{sigma(eps)}
\sigma=\varepsilon+\frac{an}{2}\varepsilon^2+\left( 9(an)^2+
 a^2- 4\dot{a}n\right)\frac{\ep^3}{24}.
\end{align}
Substituting this into  Eq.~(\ref{R(sig)}) we find:
\begin{align}\label{R(eps)}
& R^{\mu}=(n^{\mu}+v^{\mu})\varepsilon+ \left((an)v^{\mu}-
 a^{\mu}\right)\frac{ \varepsilon^2}{2}+\\ &
+\left[\left( 9(an)^2+a^2-  4\dot{a}n\right)v^{\mu}-
 12(an)a^{\mu}+ 4\dot{a}^{\mu}\right]\frac{\ep^3}{24}.\nonumber
\end{align}
Similar expansions for the velocity and the acceleration
 taken at the moment $s_{\rm{ret}}$  are
\begin{align}\label{v(eps)}
& v^{\mu}|_{s_{\rm{ret}}}=v^{\mu}-a^{\mu}\varepsilon+
\left(\dot{a}^{\mu}-(an)a^{\mu} \right)\frac{\ep^2}{2} -\\
 & - \left[\left( 9(an)^2+a^2-  4\dot{a}n\right)a^{\mu}-
 12(an)\da^{\mu}+ 4\dda^{\mu}\right]\frac{\ep^3}{24},\nonumber  \\
& a^{\mu}|_{s_{\rm{ret}}}=a^{\mu}-\dot{a}^{\mu}\varepsilon+
\left(\dda^{\mu}-(an)\dot{a}^{\mu}
 \right)\frac{\ep^2}{2}.
\end{align}
The invariant distance parameter $\rho=v_\mu(s_{\rm{ret}})R^\mu$
is given by the product of the two expansions:
\begin{align}\label{rho(eps)}
\rho=\varepsilon-an\frac{\ep^2}{2} +\left( 8(\dot{a}n)- 3 a^2-
3(an)^2\right)\frac{\ep^3}{24}.
\end{align}
For convenience we give also the expansion of the most singular
term in (\ref{Tbound4R}) up to the relevant order:
\begin{align}\label{rho6(eps)}
 \frac{1}{\rho^6}&=\frac{1}{\varepsilon^6}
\left[1+3(an)\varepsilon+ \right. \nonumber\\
& \left. +\left(6(an)^2+ 3a^2/4-2\dot{a}n\right)\varepsilon^2
\right],
\end{align}
and the expansion of the scalar product $(aR)$:
\begin{align}\label{aR(eps)}
& aR=an+\left( a^2/2-\dot{a}n\right)\varepsilon+
\left[ 4\dot{a}a/3+ a^2(an)+\right. \nonumber  \\
& \left.+ \ddot{a}v+ \ddot{a}n - (an)(\dot{a}n)\right]
\varepsilon^2/2.
\end{align}
Substituting all these expansions into  (\ref{momentvardef1}) we
obtain \begin{align}\label{N4(eps)} &\Delta
p^\mu_\bn=\frac{e^2}{4\pi}\int\limits_{s_1}^{s_2} ds
  \left\{\frac{-n^{\mu}}{2\varepsilon^2}+\frac{a^{\mu}}{2
\varepsilon}+\left[\left((an)^2+a^2/3\right)
v^{\mu}+\right.\right.\nonumber\\&+\left.\left.\left((an)^2+a^2/2\right)
n^{\mu}- 2\dot{a}^{\mu}/3  + 3(an)a^{\mu}/4\right]\right\}
 d\Omega.\end{align} Using the above rules of the
integration over the angles (\ref{sphereints}), one can see that
the leading divergent term proportional to $1/\ep^2$ vanishes and
the result reads
\begin{align}\label{M4}
\Delta p^{\mu}_\bn=e^2\int\limits_{s_1}^{s_2} ds
\left(\frac{1}{2\varepsilon} a^{\mu}-\frac{2}{3}\dot{a}^{\mu}
\right).
\end{align}
Setting $s_1=-\infty, s_2=s$, we obtain
\begin{align}\label{momentbound4}
p_\bn^{\mu} =e^2\int_{-\infty}^{s} ds'
\left(\frac{1}{2\varepsilon} a^{\mu}-\frac{2}{3}\dot{a}^{\mu}
\right).
\end{align}
Therefore the bound part of the self-force is
\begin{align}\label{boundforce}
f^{\mu}_\bn
=-\frac{dp^\mu_\bn}{ds}=-\frac{e^2a^\mu}{2\ep}+\frac{2e^2}{3}
\da^{\mu}.
\end{align}
Here the first divergent term has to be absorbed by the
renormalization of mass, while the second is the Schott term.
Comparing this with (\ref{fself}) we confirm the relation
(\ref{Schott}) and  the identity (\ref{consistency}). The Schott
term therefore is the finite part of the derivative of the bound
electromagnetic momentum of a charge. Note that {\em a priori} the
parameter of regularization $\ep$ (the radius of the small tube)
is not related to the splitting parameter   of the delta-function
in the local force calculation. But actually they give the same
form for the divergent term, for which reason we denoted them
similarly. With this convention, the divergent terms in the
momentum conservation identity (\ref{consistency}) mutually
cancel.
\section{Conclusion}
The goal of this paper was to clarify the recent discrepancy in
calculations of the bound momentum of the radiating charge and to
present an independent explicit calculation revealing the nature
of the Schott term in the Lorentz-Dirac equation. We would like to
emphasize the role of the global momentum conservation  in
understanding the origin of the Schott term. From the momentum
conservation one can derive an identity relating the contributions
to the reaction force in two alternative calculations: a local
computation on the world-line and the global integration of the
Maxwell stress-tensor. This identity demands that the Schott term
{\em should} arise from the bound field momentum, and this is
confirmed by an explicit calculation. Although this problem was
discussed by a number of authors in the past, only  few of them
presented the details of  calculation of the bound momentum, most
notably Dirac \cite{Di38}, with whom we  perfectly agree. We
disagree, however, with more recent results of Poisson
\cite{Po99}, who used a different method, based on retarded
coordinated, in order to avoid somewhat lengthy expansions used in
the Dirac's calculation. According to \cite{Po99}, the bound
momentum reduces entirely to the mass-renormalization term. As we
have shown here, this result is incompatible with the momentum
conservation identity and thus can not be correct. More detailed
analysis shows that the integration surface in \cite{Po99} does
not correspond to the definition of the bound momentum associated
with the given moment of the proper time of a charge (see the
Appendix below). Therefore we confirm the result of Dirac and show
that no alternative interpretation of the Schott term is needed
(and not possible, moreover). Technically, our derivation is
slightly different from that of Dirac in performing all the
necessary expansions in a uniform way. Conceptually, we fully
agree with Teitelboim~\cite{Te70}, presenting in addition full
details of the global derivation of the Schott term.

Notice that in the literature one often encounters a confusing
notation related to two alternative ways to select radiation.
Namely, one has to distinguish the 'radiative' field as the
half-difference of the retarded and advanced fields and the
'radiative' component of the Maxwell stress-tensor, which is
defined through the retarded field. Here we suggest a different
wording for the  radiative part of the stress tensor, which we
call 'emitted'. This distinction is necessary indeed since the
corresponding local forces differ by the Schott term.

Physically, the Schott term describes the reversible variation of
the Coulomb field bound momentum during an accelerated motion of
the charge. Variation of the {\em mechanical} momentum of the
charge consists of this reversible part (which may be both
positive or negative) and an irreversible loss due to radiation.
In other words, the momentum radiated away can be borrowed both
from the mechanical momentum and from the bound Coulomb momentum,
and this explains how in the threshold case of the uniform
acceleration the mechanical momentum of the radiating charge may
remain constant. One is not allowed to redefine the  mechanical
momentum of the charge by adding to it the Schott term without
facing contradiction with the global energy-momentum conservation
identity.

\begin{acknowledgments}
D.V.G. is grateful to the University of Tours for hospitality at
some stage of this work. He would like to thank P. Forgacs and M.
Volkov for useful discussions. The work was supported in part by
the RFBR grant 02-04-16949.
\end{acknowledgments}
\subsection{Appendix. Integration of flux in retarded coordinates}
Here for comparison we reproduce the formulation of the problem in
the retarded coordinates~\cite{Po99} (recently generalized to
arbitrary dimensions in~\cite{Ya03}). The main difference with our
procedure is the definition of the integration hypersurface for
the Maxwell stress-tensor in terms of the retarded coordinates due
to Newman and Unti \cite{NeUn63} $(s=s_{\rm ret},\rho,\theta_i )$.
These are introduced as follows. Selecting an arbitrary point
$z(s)$ on the world-line one constructs its future light cone
(Fig. 4)
\begin{figure}
\centerline{\psfrag{s}{\LARGE $ s$} \psfrag{ser}{\LARGE $
\hat{s}$} \psfrag{Hser}{\LARGE $ H(s,\rho)$} \psfrag{Cser}{\LARGE
$ C(s)$} \psfrag{Sser}{\LARGE $
\Sigma(s,\rho)$}\psfrag{zmis}{\LARGE $
z^{\mu}(s)$}\psfrag{zmis}{\LARGE $ z^{\mu}(s)$}
\psfrag{Sro}{\LARGE $S_{\rho}$}
 \includegraphics
[angle=270,width=14cm]{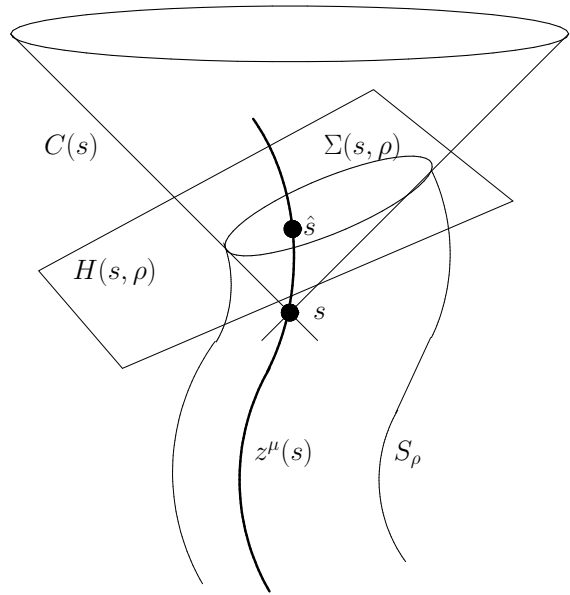}}\caption{Integration tube in the
retarded coordinates. Here $C(s)$ is a future light cone with an
apex at  $z(s)$,\; $H(s,\rho)$ is a hyperplane transverse to
$v(s)$ which intersects the world-line at a point $ \hat{s}$. The
two-dimensional surface $ \Sigma(s,\rho)$ is an intersection
$\Sigma(s,\rho)=C(s)\bigcap H(s,\rho),$ $S_{\rho}$ is a lateral
hypersurface of constant $\rho,$ formed by all $\Sigma(s',\rho)$,
$s'=-\infty...s.$} \label{figur4}
\end{figure}
\be\lb{C}C(s)=\{x|(x-z(s))^2=0, x^0\geqslant z^0\},\ee ascribing
to it  the unique value of the coordinate $s$,  and defines on it
the coordinate $\rho$ as an affine length parameter on a null
geodesic specified by the angle coordinates $\theta_i$. The  orbit
of the constant $\rho(s,x)$ on the cone (\ref{C}) forms a
two-dimensional manifold $\Sigma(s,\rho)$ which is an intersection
of the hyperplane \be\lb{H}H(s,\rho)=\{x|(x-z(s))^\mu
v_\mu(s)=\rho=\rm const\}\ee and the cone (\ref{C}).  The  open
tube surrounding the world-line is then defined as the sequence
$\Sigma(u,\rho)$ of hypersurfaces $\rho=\mathrm{const}$ for $u\in
(-\infty,s]$. The area element on this tube is equal to \be
\lb{dSP}dS_{\mu}=\rho^{2}(v_{\mu}(u)+\lambda(u,x) c_{\mu}(u,x))du
d\Omega,\ee where   $c_\mu=(x-z(u))_\mu/\rho$. The factor $\rho$
here is the same as in the expressions (\ref{SET4Dradiat}),
(\ref{Tbound4c}), so it  partially compensates the denominator of
$T^{\mu\nu}_\bn$ and fully compensates that of $T^{\mu\nu}_\emi$,
which looks as  simplification. The procedure proposed in
\cite{Po99} consists in the integration of the flux over the tube
for fixed $\rho$ with the subsequent limit $\rho\to 0$.

However, this procedure has a serious drawback when applied to the
bound part of the electromagnetic momentum because of the singular
nature of the integrand at $\rho=0$. In fact, this means that we
have to consider the sequence of tubes of variable radius $\rho$.
But with variable $\rho$   the integral
\begin{align}\label{poissonint}
 \int \limits_{-\infty}^s du \int  \limits  T^{\mu
 \nu} \rho^{ 2}(v_{\mu}+(\rho(ac)-1)c_{\mu})  d\Omega
\end{align}
does not give the field momentum associated with any given moment
of the proper time. Indeed, when $\rho$ is changing, the spheres
$\Sigma(s,\rho)$ move across the light cone and do not lie on a
definite space-like hypersurface.   For any fixed finite $\rho$
the integral (\ref{poissonint}) is performed over the hyperplane
intersecting the world-line at the point of proper time
\be\lb{hats}\hat{s}=s+\tau (s,\rho)>s,\ee as shown on Fig. 3. Only
in the limit ${\hat s}\to s$ one actually integrates over the
hyperplane intersecting the world-line at $s$,  but performing the
limit as $\rho\to 0$ one passes through the sequence of {\em
different} space-like hyperplanes.  Contrary to this, our
procedure consists in fixing the unique space-like hyperplane and
carefully expanding the integrand in powers of the radius of the
tube. This allows to pass consistently to the limit of the
vanishing tube radius.

Therefore it is not surprising that a calculation along the lines
of \cite{Po99} gives for the bound momentum only the leading
divergent term  (see the Eq. (8.3) of \cite{Po99})\be
\frac{dp^\mu_\bn}{ds}=\frac{e^2a^\mu}{2\ep},\ee but fails to
produce the second finite term in (\ref{boundforce}). For the
emitted momentum the procedure of \cite{Po99} works, since the
answer in this case is given entirely  by the leading term.

\end{document}